# Involutive Categories and Monoids, with a GNS-correspondence

Bart Jacobs

*Institute for Computing and Information Sciences (iCIS)*
*Radboud University Nijmegen, The Netherlands*

August 2, 2018

**Abstract**

This paper develops the basics of the theory of involutive categories and shows that such categories provide the natural setting in which to describe involutive monoids. It is shown how categories of Eilenberg-Moore algebras of involutive monads are involutive, with conjugation for modules and vector spaces as special case. The core of the so-called Gelfand-Naimark-Segal (GNS) construction is identified as a bijective correspondence between states on involutive monoids and inner products. This correspondence exists in arbitrary involutive (symmetric monoidal) categories.

## 1 Introduction

In general an involution is a certain endomap $i$ for which $i \circ i$ is the identity. The inverse operation of a group is a special example. But there are also monoids with such an involution, such as for instance the free monoid of lists, with list reversal as involution.

An involution can also be defined on a category. It then consists of an endofunctor $\mathbf{C} \to \mathbf{C}$, which is typically written as $X \mapsto \overline{X}$. It should satisfy $\overline{\overline{X}} \cong X$. Involutive categories occur in the literature, for instance in [1], but have, as far as we know, not been studied systematically. Involutions are of particular importance in the (categorical) foundations of quantum mechanics and computing, see [2]. This paper will develop the basic elements of such a theory of involutive categories.

We should note that involutive categories as we understand them here are different from dagger categories (which have an identity-on-objects functor $(-)^{\dagger} \colon \mathbf{C}^{\mathrm{op}} \to \mathbf{C}$ with $f^{\dagger\dagger} = f$) and also from $*$-autonomous categories (which have a duality $(-)^{*} \colon \mathbf{C}^{\mathrm{op}} \to \mathbf{C}$ given by a dualising object $D$ as in $X^{*} = X \multimap D$). In both these cases one has contravariant functors, whereas involution $\overline{(-)} \colon \mathbf{C} \to \mathbf{C}$ is a covariant functor. The relation between involution, dagger and duality for Hilbert spaces is described in [2, §§4.1, 4.2]: each can be defined in terms of the other two.

Involutive categories and involutive monoids are related: just like the notion of a monoid is formulated in a monoidal category, the notion of involutive monoid requires an appropriate notion of involutive monoidal category. This is in line with the "microcosm principle", formulated by Baez and Dolan [4], and elaborated in [12,11,10]: it involves "outer" structure (like monoidal structure $1 \xrightarrow{I} \mathbf{C} \xleftarrow{\otimes} \mathbf{C} \times \mathbf{C}$ *on* a category $\mathbf{C}$) that enables the definition of "inner" structure (like a monoid $I \xrightarrow{0} M \xleftarrow{+} M \otimes M$ *in* $\mathbf{C}$). We briefly illustrate how this connection between involutive monoids and involutive categories arises.

Consider for instance the additive group $\mathbb{Z}$ of integers with minus $-$ as involution. In the category **Sets** of ordinary sets and functions between them we can describe minus as an ordinary endomap $-\colon \mathbb{Z} \to \mathbb{Z}$. The integers form a partially ordered set, so we may wish to consider $\mathbb{Z}$ also as involutive monoid in the category **PoSets** of partially ordered sets and monotone functions. The problem is that minus reverses the order: $i \leq j \Rightarrow -i \geq -j$, and is thus not a map $\mathbb{Z} \to \mathbb{Z}$ in **PoSets**. However, we can describe it as a map $(\mathbb{Z}, \geq) \to (\mathbb{Z}, \leq)$ in **PoSets**, using the reversed order ($\geq$ instead of $\leq$) on the integers. This order reversal forms an involution $\overline{(-)}\colon \mathbf{PoSets} \to \mathbf{PoSets}$ on the "outer" category, which allows us to describe the involution "internally" as $-\colon \overline{\mathbb{Z}} \to \mathbb{Z}$ in **PoSets**.

As said, this paper introduces the basic steps of the theory of involutive categories. It introduces the category of "self-conjugate" objects, and shows how involutions arise on categories of Eilenberg-Moore algebras of an "involutive" monad. This general construction includes the important example of conjugation on modules and vector spaces, for the multiset monad associated with an involutive semiring. It allows us to describe abstractly an involutive monoid in such categories of algebras. Pre $C^*$-algebras (without norm) are such monoids.

Once this setting has been established we take a special look at the famous Gelfand-Naimark-Segal (GNS) construction [3]. It relates $C^*$-algebras and Hilbert spaces, and shows in particular how a state $A \to \mathbb{C}$ on a $C^*$-algebra gives rise an inner product on $A$. Using conjugation as involution, the latter can be described as a map $\overline{A} \otimes A \to \mathbb{C}$ that incorporates the sesquilinearity requirements in its type (including conjugate linearity in its first argument). The final section of this paper gives the essence of this construction in the form of a bijective correspondence between such states and inner products in categorical terms, using the language of involutive categories and monoids.

## 2 Involutive categories

This section only contains the most basic material.

**Definition 2.1** A category $\mathbf{C}$ will be called *involutive* if it comes with a functor $\mathbf{C} \to \mathbf{C}$, written as $X \mapsto \overline{X}$, and a natural isomorphism $\iota\colon X \xrightarrow{\cong} \overline{\overline{X}}$ satisfying

$$\begin{array}{ccc} \overline{X} & \xrightarrow{\iota_{\overline{X}}} & \overline{\overline{\overline{X}}} \\ \| & & \| \\ \overline{X} & \xrightarrow{\overline{\iota_X}} & \overline{\overline{\overline{X}}} \end{array} \qquad (1)$$

Each category is trivially involutive via the identity functor. This trivial involu-tion is certainly useful. The category **PoSets** is involutive via order reversal. This applies also to categories of, for instance, distributive lattices or Boolean algebras. The category **Cat** of (small) categories and functors is also involutive, by taking opposites of categories. Next, consider the category $\mathit{Vect}_{\mathbb{C}}$ of vector spaces over the complex numbers $\mathbb{C}$. It is an involutive category via conjugation (see Example 6.5 later on, for a systematic description). For a vector space $V \in \mathit{Vect}_{\mathbb{C}}$ we define $\overline{V} \in \mathit{Vect}_{\mathbb{C}}$ with the same vectors as $V$, but with adapted scalar multiplication: for $s \in \mathbb{C}$ and $v \in V$,

$$s \cdot_{\overline{V}} v = \overline{s} \cdot_V v, \tag{2}$$

where $\overline{s} = a - ib$ is the conjugate of the complex number $s = a + ib \in \mathbb{C}$.

The following is the first of a series of basic observations.

**Lemma 2.2** *The involution functor of an involutive category is self-adjoint:* $\overline{(-)} \dashv \overline{(-)}$. *As a result, involution preserves all limits and colimts that exist in the category.*

**Proof.** Obviously there are bijective correspondences:

$$\frac{\overline{X} \xrightarrow{f} Y}{\overline{\overline{X \xrightarrow[g]{} \overline{Y}}}}$$

One maps $f$ to $\widehat{f} = \overline{f} \circ \iota_X \colon X \xrightarrow{\cong} \overline{\overline{X}} \to \overline{Y}$ and $g$ to $\widehat{g} = \iota_Y^{-1} \circ \overline{g} \colon \overline{X} \to \overline{\overline{Y}} \xrightarrow{\cong} Y$. □

**Definition 2.3** A functor $F \colon \mathbf{C} \to \mathbf{D}$ between two involutive categories is called involutive if it comes with a natural transformation (or distributive law) $\nu$ with components $F(\overline{X}) \to \overline{F(X)}$ commuting appropriately with the isomorphisms $X \cong \overline{\overline{X}}$, as in:

$$\begin{array}{ccccc} F(X) & = & = & = & F(X) \\ {\scriptstyle F(\iota_X)}\downarrow\cong & & & & \cong\downarrow{\scriptstyle \iota_{F(X)}} \\ F(\overline{\overline{X}}) & \xrightarrow[\nu_{\overline{X}}]{} & \overline{F(\overline{X})} & \xrightarrow[\overline{\nu_X}]{} & \overline{\overline{F(X)}} \end{array} \tag{3}$$

A natural transformation $\sigma \colon F \Rightarrow G$ between two involutive functors $F, G \colon \mathbf{C} \rightrightarrows \mathbf{D}$ is called involutive if it commutes with the associated $\nu$'s, as in:

$$\begin{array}{ccc} F(\overline{X}) & \xrightarrow{\sigma_{\overline{X}}} & G(\overline{X}) \\ {\scriptstyle \nu^F}\downarrow & & \downarrow{\scriptstyle \nu^G} \\ \overline{F(X)} & \xrightarrow{\overline{\sigma_X}} & \overline{G(X)} \end{array}$$

In this way we obtain a 2-category **ICat** of involutive categories, functors and natural transformations.

This 2-categorical perspective is useful, for instance because it allows us to see immediately what an involutive adjunction or monad is, namely one in which the functors and natural transformations involved are all involutive.

**Lemma 2.4** *If $F$ is an involutive functor via $\nu \colon F(\overline{X}) \to \overline{F(X)}$, then this $\nu$ is automatically an isomorphism.*

**Proof.** We construct an inverse for $\nu$ as composite:

$$\nu^{-1} \stackrel{\text{def}}{=} \Big(\overline{F(X)} \xrightarrow{\overline{F(\iota)}} \overline{F(\overline{\overline{X}})} \xrightarrow{\overline{\nu}} \overline{\overline{F(\overline{X})}} \xrightarrow{\iota^{-1}} F(\overline{X})\Big).$$

We explicitly check that this is indeed an inverse to $\nu$, by using the interaction (3) between $\nu$ and $\iota$. First we have $\nu \circ \nu^{-1} = \mathrm{id}$ in:

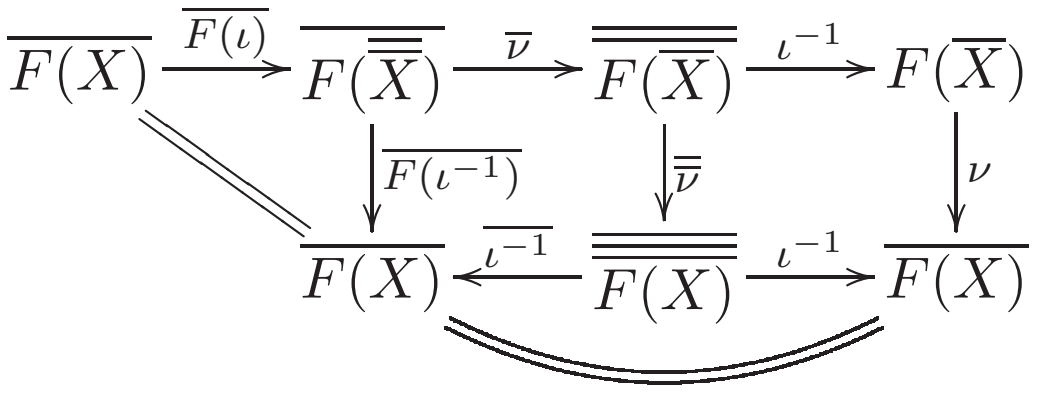

And similarly we get $\nu^{-1} \circ \nu = \mathrm{id}$ in:

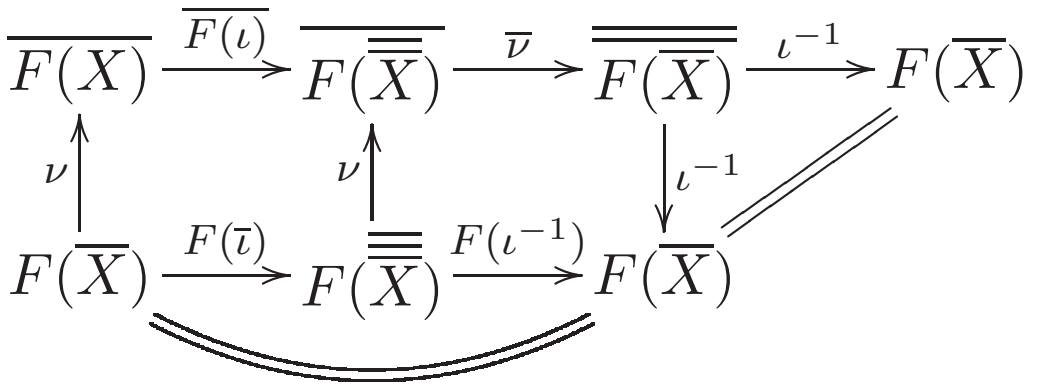

□

# 3 Self-conjugates

**Definition 3.1** For an involutive category $\mathbf{C}$, let $SC(\mathbf{C})$ be the category of self-conjugates in $\mathbf{C}$. Its objects are maps $j \colon \overline{X} \to X$ making the triangle below commute.

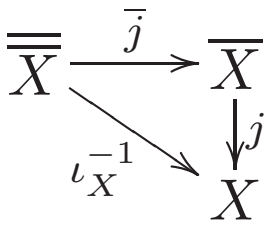

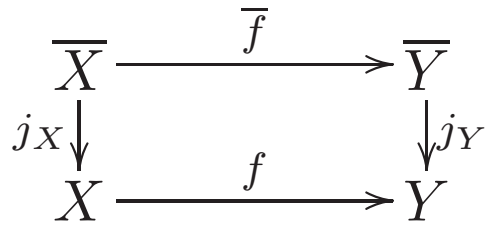

It is not hard to see that such a map is $j$ is necessarily an isomorphism, with inverse $\overline{j} \circ \iota_X \colon X \to \overline{\overline{X}} \to \overline{X}$.

A morphism $f \colon (X, j_X) \to (Y, j_Y)$ in $SC(\mathbf{C})$ is a map $f \colon X \to Y$ in $\mathbf{C}$ making the above rectangle commute. There is thus an obvious forgetful functor $SC(\mathbf{C}) \to \mathbf{C}$.

By the self-adjointness of Lemma 2.2 a self-conjugate $\overline{X} \to X$ may also be described as $X \to \overline{X}$. Sometimes we call an object $X$ a self-conjugate when the map $\overline{X} \xrightarrow{\cong} X$ involved is obvious from the context. In linear algebra, with $\overline{X}$ given by conjugation (see (2), or also Example 6.5), a map of the form $\overline{X} \to Y$ is called an 'antilinear' or 'conjugate linear' map.

Before describing examples we first note the following. A more systematic description follows in Lemma 3.6.

**Lemma 3.2** *For an involutive category* $\mathbf{C}$*, the category* $SC(\mathbf{C})$ *of self-conjugates is again involutive, via:*

$$\overline{\big(\overline{X} \xrightarrow{j} X\big)} \stackrel{\text{def}}{=} \big(\overline{\overline{X}} \xrightarrow{\overline{j}} \overline{X}\big). \tag{4}$$

*and the forgetful functor $SC(\mathbf{C}) \to \mathbf{C}$ is an involutive functor, via the identity natural transformation (as '$\nu$' in Definition 2.3).*

**Proof.** The map $\iota_X : X \xrightarrow{\cong} \overline{\overline{X}}$ in $\mathbf{C}$ is also a map in $SC(\mathbf{C})$ in:

$$\big(\overline{X} \xrightarrow{j} X\big) \xrightarrow[\cong]{\iota_X} \overline{\overline{\big(\overline{X} \xrightarrow{j} X\big)}}$$

since the following diagram commutes by naturality.

$$\begin{array}{ccc} \overline{X} & \xrightarrow{\overline{\iota_X} = \iota_{\overline{X}}} & \overline{\overline{\overline{X}}} \\ {\scriptstyle j}\downarrow & & \downarrow{\scriptstyle \overline{\overline{j}}} \\ X & \xrightarrow{\iota_X} & \overline{\overline{X}} \end{array}$$

□

**Example 3.3** Recall that the category **PoSets** of posets and monotone functions is involutive via the reversed (opposite) order: $\overline{(X, \leq)} = (X, \geq)$. The integers $\mathbb{Z}$ are then self-conjugate, via minus $- : \overline{\mathbb{Z}} \xrightarrow{\cong} \mathbb{Z}$. Also the positive rational and real numbers $\mathbb{Q}_{>0}$ and $\mathbb{R}_{>0}$ are self-conjugates in **PoSets**, via $x \mapsto \frac{1}{x}$. Similarly, for a Boolean algebra $B$, negation $\neg$ yields a self-conjugate $\neg : \overline{B} \xrightarrow{\cong} B$ in the category of Boolean algebras. There are similar self-conjugates via orthosupplements $(-)^{\perp}$ in orthomodular lattices [13] and effect algebras [9].

In **Cat** a self-conjugate is given by a self-dual category $\mathbf{C}^{\mathrm{op}} \cong \mathbf{C}$.

Recall the conjugation (2) on vector spaces. Suppose $V \in \mathit{Vect}_{\mathbb{C}}$ has a basis $(v_i)_{i \in I}$. Then we can define a self-conjugate $\overline{V} \xrightarrow{\cong} V$ by:

$$x = \big(\textstyle\sum_i x_i v_i\big) \longmapsto \big(\textstyle\sum_i \overline{x_i} v_i\big).$$

Finally, if a category $\mathbf{C}$ is considered with trivial involution $\overline{X} = X$, then $SC(\mathbf{C})$ contains the self-inverse endomaps $j : X \to X$, with $j \circ j = \mathrm{id}_X$.

We first take a closer look at these trivial involutions.

**Lemma 3.4** *Let $\mathbf{C}$ be an ordinary category, considered as involutive with trivial involution $\overline{X} = X$. Assuming binary coproducts $+$ and products $\times$ exist in $\mathbf{C}$, there are left and right adjoints to the forgetful functor:*

$$\begin{array}{c} SC(\mathbf{C}) \\ {\scriptstyle X \mapsto 2 \times X = X + X}\ \big(\ \dashv\ \downarrow\ \dashv\ \big)\ {\scriptstyle X \mapsto X^2 = X \times X} \\ \mathbf{C} \end{array}$$

*using the swap maps $[\kappa_2, \kappa_1] : X + X \xrightarrow{\cong} X + X$ and $\langle \pi_2, \pi_1 \rangle : X \times X \xrightarrow{\cong} X \times X$ as self-conjugates.*

**Proof.** Recall that for the trivial involution on $\mathbf{C}$, an object $(Y, j) \in SC(\mathbf{C})$ consists of an isomorphism $j : Y \xrightarrow{\cong} Y$ with $j^{-1} = j$. For the left adjoint the required bijective correspondence:

$$\begin{array}{cc} (X + X, [\kappa_2, \kappa_1]) \xrightarrow{f} (Y, j) & \text{in } SC(\mathbf{C}) \\ \hline\hline X \xrightarrow[g]{} Y & \text{in } \mathbf{C} \end{array}$$

exists because the requirement $f \circ [\kappa_2, \kappa_1] = j \circ f$ means $f \circ \kappa_2 = j \circ f \circ \kappa_1$. Hence $f$ is determined by $f \circ \kappa_1 \colon X \to Y$. The argument works similarly for the right adjoints, given by products. □

**Lemma 3.5** *Let* $\mathbf{C}$ *be an involutive category;* $SC(\mathbf{C})$ *inherits all limits and colimits that exist in* $\mathbf{C}$*, and the forgetful functor* $SC(\mathbf{C}) \to \mathbf{C}$ *preserves them.*

**Proof.** We give an exemplaric sketch for binary products $\times$. The product of two objects $(X, j_X), (Y, j_Y) \in SC(\mathbf{C})$ is given by:

$$\overline{X \times Y} \xrightarrow[\cong]{\langle \overline{\pi_1}, \overline{\pi_2} \rangle} \overline{X} \times \overline{Y} \xrightarrow{j_X \times j_Y} X \times Y,$$

where the (canonical) isomorphism exists since $\overline{(-)}$ preserves products, by Lemma 2.2. It is not hard to see that this is a self-conjugate, forming a product in $SC(\mathbf{C})$. □

For the record we note the following (see [18,6] for background).

**Lemma 3.6** *The mapping* $\mathbf{C} \mapsto SC(\mathbf{C})$ *is a 2-functor* $\mathbf{ICat} \to \mathbf{ICat}$*, and even a 2-comonad.*

**Proof.** Essentially this says that we can lift involutive functors and natural transformations as in:

$$\begin{array}{ccc} SC(\mathbf{C}) & \underset{SC(G)}{\overset{SC(F)}{\rightrightarrows}} \; \Downarrow SC(\sigma) & SC(\mathbf{D}) \\ \downarrow & & \downarrow \\ \mathbf{C} & \underset{G}{\overset{F}{\rightrightarrows}} \; \Downarrow \sigma & \mathbf{D} \end{array} \tag{5}$$

Using Lemma 2.4 the lifted functor $SC(F)$ is defined as:

$$\Big(\overline{X} \xrightarrow{j} X\Big) \longmapsto \Big(\overline{F(X)} \xrightarrow{\nu^{-1}} F(\overline{X}) \xrightarrow{F(j)} F(X)\Big). \tag{6}$$

It is not hard to see that the right-hand-side is a again a self-conjugate. The natural transformation $SC(\sigma)$ on $\overline{X} \to X$ is simply $\sigma_X$.

The counit of $SC$ as 2-comonad is the forgetful functor $SC(\mathbf{C}) \to \mathbf{C}$, which is natural, see (5). The comultiplication $SC(\mathbf{C}) \to SC(SC(\mathbf{C}))$ is given by:

$$(\overline{X} \xrightarrow{j} X) \longmapsto \Big(\overline{(\overline{X} \xrightarrow{j} X)} \xrightarrow{j} (\overline{X} \xrightarrow{j} X)\Big). \qquad \square$$

# 4 Involutive monoidal categories

**Definition 4.1** An *involutive monoidal category* or an *involutive symmetric monoidal category*, abbreviated as IMC or ISMC, is a category $\mathbf{C}$ which is both involutive and (symmetric) monodial in which involution $\overline{(-)} \colon \mathbf{C} \to \mathbf{C}$ is a (symmetric) monoidal functor and $\iota \colon \mathrm{id} \Rightarrow \overline{\overline{(-)}}$ is a monoidal natural transformation.

The fact that involution is a (symmetric) monoidal functor means that there are (natural) maps $\zeta\colon I \to \overline{I}$ and $\xi\colon \overline{X} \otimes \overline{Y} \to \overline{X \otimes Y}$ commuting with the monoidal isomorphisms $\alpha\colon X \otimes (Y \otimes Z) \overset{\cong}{\Rightarrow} (X \otimes Y) \otimes Z$, $\lambda\colon I \otimes X \overset{\cong}{\Rightarrow} X$, $\rho\colon X \otimes I \to X$, and also with the swap map $\gamma\colon X \otimes Y \overset{\cong}{\Rightarrow} Y \otimes X$ in the symmetric case. That the isomorphism $\iota$ is monoidal means that we have commuting diagrams:

$$\begin{array}{ccccc} I & = & = & = & I \\ \| & & & & \downarrow{\scriptstyle\iota} \\ I & \xrightarrow{\zeta} & \overline{I} & \xrightarrow{\overline{\zeta}} & \overline{\overline{I}} \end{array} \qquad \begin{array}{ccccc} X \otimes Y & = & = & = & X \otimes Y \\ {\scriptstyle\iota\otimes\iota}\downarrow & & & & \downarrow{\scriptstyle\iota} \\ \overline{\overline{X}} \otimes \overline{\overline{Y}} & \xrightarrow{\xi} & \overline{\overline{X} \otimes \overline{Y}} & \xrightarrow{\overline{\xi}} & \overline{\overline{X \otimes Y}} \end{array} \tag{7}$$

Like in Lemma 2.4 we get isomorphy for free.

**Lemma 4.2** *In an IMC the involution functor $\overline{(-)}$ is automatically strong monoidal: the maps $\zeta\colon I \to \overline{I}$ and $\xi\colon \overline{X} \otimes \overline{Y} \to \overline{X \otimes Y}$ are necessarily isomorphisms.*

**Proof.** All this follows from the requirement $\iota = \overline{\iota}\colon \overline{X} \to \overline{\overline{\overline{X}}}$ in (1) in Definition 2.1 and the monoidal requirements (7). For instance, the obvious candidate as inverse for $\zeta\colon I \to \overline{I}$ is $\iota^{-1} \circ \overline{\zeta}\colon \overline{I} \to \overline{\overline{I}} \overset{\cong}{\Rightarrow} I$. Because $\iota$ is a monoidal natural transformation, we immediately get $\iota^{-1} \circ \overline{\zeta} \circ \zeta = \iota^{-1} \circ \iota = \mathrm{id}$. By post-composing with the isomorphism $\iota = \overline{\iota}\colon \overline{I} \to \overline{\overline{\overline{I}}}$ we get by (7):

$$\iota \circ \zeta \circ \iota^{-1} \circ \overline{\zeta} = \overline{\overline{\zeta}} \circ \iota \circ \iota^{-1} \circ \overline{\zeta} = \overline{\overline{\zeta} \circ \zeta} = \overline{\iota} = \iota.$$

Similarly, the (candidate) inverse for $\xi\colon \overline{X} \otimes \overline{Y} \to \overline{X \otimes Y}$ is:

$$\overline{X \otimes Y} \xrightarrow{\iota\otimes\iota} \overline{\overline{\overline{X}} \otimes \overline{\overline{Y}}} \xrightarrow{\overline{\xi}} \overline{\overline{\overline{X} \otimes \overline{Y}}} \xrightarrow{\iota^{-1}} \overline{X} \otimes \overline{Y}. \qquad \square$$

In order to be complete we also have to define the following.

**Definition 4.3** A functor $F\colon \mathbf{C} \to \mathbf{D}$ between IMC's is called *involutive monoidal* if it is both involutive, via $\nu\colon F(\overline{X}) \to \overline{F(X)}$, and monoidal, via $\zeta^F\colon I \to F(I)$ and $\xi^F\colon F(X) \otimes F(Y) \to F(X \otimes Y)$, and these natural transformations $\nu, \zeta^F, \xi^F$ interact appropriately with $\zeta, \xi$ from(7), as in:

$$\begin{array}{ccccc} I & \xrightarrow{\zeta^F} & F(I) & \xrightarrow{F(\zeta)} & F(\overline{I}) \\ \| & & & & \downarrow{\scriptstyle\nu} \\ I & \xrightarrow{\zeta} & \overline{I} & \xrightarrow{\overline{\zeta^F}} & \overline{F(I)} \end{array} \qquad \begin{array}{ccccc} F(\overline{X}) \otimes F(\overline{Y}) & \xrightarrow{\xi^F} & F(\overline{X} \otimes \overline{Y}) & \xrightarrow{F(\xi)} & F(\overline{X \otimes Y}) \\ {\scriptstyle\nu\otimes\nu}\downarrow & & & & \downarrow{\scriptstyle\nu} \\ \overline{F(X)} \otimes \overline{F(Y)} & \xrightarrow{\xi} & \overline{F(X) \otimes F(Y)} & \xrightarrow{\overline{\xi^F}} & \overline{F(X \otimes Y)} \end{array}$$

It should then be obvious what an involutive *symmetric* monoidal functor is.

An involutive monoidal natural transformation $\sigma\colon F \Rightarrow G$ between two involutive monoidal functors is both involutive and monoidal.

Hence also in this case we have two categories **IMCat** and **IMSCat** of involutive (symmetric) monoidal categories.

Now we come to the main result of this section.

**Proposition 4.4** *A category $SC(\mathbf{C})$ inherits (symmetric) monoidal structure from $\mathbf{C}$. As a result, the forgetful functor $SC(\mathbf{C}) \to \mathbf{C}$ is an involutive (symmetric) monoidal functor.*

*In case* $\mathbf{C}$ *is monoidal closed, then so is* $SC(\mathbf{C})$ *and* $SC(\mathbf{C}) \to \mathbf{C}$ *preserves the exponent* $\multimap$*.*

**Proof.** The tensor unit $I \in \mathbf{C}$ is a self-conjugate via $\zeta^{-1} \colon \overline{I} \xrightarrow{\cong} I$. If we have self-conjugates $j_X \colon \overline{X} \xrightarrow{\cong} X$ and $j_Y \colon \overline{Y} \xrightarrow{\cong} Y$ we obtain a tensored self-conjugate using Lemma 4.2:

$$\overline{X \otimes Y} \xrightarrow[\cong]{\xi^{-1}} \overline{X} \otimes \overline{Y} \xrightarrow[\cong]{j_X \otimes j_Y} X \otimes Y.$$

It is not hard to see that, with this tensor product, the monoidal isomorphisms $\alpha, \lambda, \rho, \gamma$ from $\mathbf{C}$ are also maps in $SC(\mathbf{C})$. Similarly, for the required maps making the involution $\overline{(-)} \colon SC(\mathbf{C}) \to SC(\mathbf{C})$ from Lemma 3.2 into a monoidal functor, we can take the ones from $\mathbf{C}$, in:

$$\big(\overline{I} \xrightarrow[\zeta^{-1}]{} I\big) \xrightarrow{\zeta} \overline{\big(\overline{I} \xrightarrow[\zeta^{-1}]{} I\big)} \qquad \overline{\big(\overline{X} \xrightarrow[j_X]{} X\big)} \otimes \overline{\big(\overline{Y} \xrightarrow[j_Y]{} Y\big)} \xrightarrow{\xi} \overline{\big(\overline{X \otimes Y} \xrightarrow[j_X \otimes j_Y \circ \xi^{-1}]{} X \otimes Y\big)},$$

The exponent of $(X, j_X), (Y, j_Y) \in SC(\mathbf{C})$ is $X \multimap Y$ with self-conjugate $\overline{X \multimap Y} \to X \multimap Y$ obtained by abstraction from:

$$X \otimes \overline{(X \multimap Y)} \xrightarrow{j_X^{-1} \otimes \mathrm{id}} \overline{X} \otimes \overline{(X \multimap Y)} \xrightarrow{\xi} \overline{X \otimes (X \multimap Y)} \xrightarrow{\overline{\mathrm{ev}}} \overline{Y} \xrightarrow{j_Y} Y. \qquad \square$$

## 5 Involutive Monoids

Now that we have the notion of involutive category as ambient category, we can define the notion of involutive monoid in this setting, in the style of [12,11,10].

We start with some preliminary observations. Let $M = (M, \cdot, 1)$ be an arbitrary monoid (in **Sets**), not necessarily commutative. An involution on $M$ is a special endofunction $M \to M$ which we shall write as superscript negation $x^-$, for $x \in M$. It satisfies $x^{--} = x$ and $1^- = 1$. The interaction of involution and multiplication may happen in two ways: either in a "reversing" manner, as in $(x \cdot y)^- = y^- \cdot x^-$, or in a "non-reversing" manner: $(x \cdot y)^- = x^- \cdot y^-$. Obviously, in a commutative monoid there is no difference between a reversing or non-reversing involution.

Each group is a reversing involutive monoid with $x^- = x^{-1}$. One advantage of involutive monoids over groups is that they involve only "linear" equations, with axioms containing variables exactly once on both sides of the equation sign. Groups however are non-linear, via the axiom $x \cdot x^{-1} = 1$. Hence this equation cannot be formulated in a monoidal category, since it requires diagonals and projections. Instead, one commonly uses Hopf algebras.

As we have argued in the first section via the example of integers in **PoSets**, a proper formulation of the notion of involutive monoid requires an involutive category, so that the monoid involution can be described as a map $\overline{M} \to M$.

**Definition 5.1** Let $\mathbf{C}$ be an involutive symmetric monoidal category. An involutive monoid in $\mathbf{C}$ consists of a monoid $I \xrightarrow{u} M \xleftarrow{m} M \otimes M$ in $\mathbf{C}$ together with an involution map $\overline{M} \stackrel{j}{\to} M$ satisfying:

$$\begin{array}{ccc} I & \xrightarrow{u} & M \\ {\scriptstyle\zeta}\downarrow{\scriptstyle\cong} & & \uparrow{\scriptstyle j} \\ \overline{I} & \xrightarrow{\overline{u}} & \overline{M} \end{array} \qquad\qquad \begin{array}{ccc} \overline{\overline{M}} & \xrightarrow{\overline{j}} & \overline{M} \\ & {\scriptstyle\iota^{-1}}\searrow{\scriptstyle\cong} & \downarrow{\scriptstyle j} \\ & & M \end{array}$$

and, one of the following diagrams:

"reversing"

$$\begin{array}{ccccc} \overline{M} \otimes \overline{M} & \xrightarrow{\xi} & \overline{M \otimes M} & \xrightarrow{\overline{m}} & \overline{M} \\ {\scriptstyle j\otimes j}\downarrow & & & & \downarrow{\scriptstyle j} \\ M \otimes M & \xrightarrow[\cong]{\gamma} & M \otimes M & \xrightarrow{m} & M \end{array}$$

"non-reversing"

$$\begin{array}{ccccc} \overline{M} \otimes \overline{M} & \xrightarrow{\xi} & \overline{M \otimes M} & \xrightarrow{\overline{m}} & \overline{M} \\ {\scriptstyle j\otimes j}\downarrow & & & & \downarrow{\scriptstyle j} \\ M \otimes M & & \xrightarrow{\qquad m \qquad} & & M \end{array}$$

One may call $M$ a *simple* involutive monoid if $\mathbf{C}$'s involution $\overline{(-)}$ is the identity.

A morphism of involutive monoids $M \to M'$ is a morphism of monoids $f\colon M \to M'$ satisfying $f \circ j = j' \circ \overline{f}$. This yields two subcategories $\mathbf{rIMon}(\mathbf{C}) \hookrightarrow \mathbf{Mon}(\mathbf{C})$ and $\mathbf{IMon}(\mathbf{C}) \hookrightarrow \mathbf{Mon}(\mathbf{C})$ of reversing and non-reversing involutive monoids. There is also a commutative version, forming a (full) subcategory. $\mathbf{ICMon}(\mathbf{C}) \hookrightarrow \mathbf{IMon}(\mathbf{C})$.

The involution map $j\colon \overline{M} \to M$ of an involutive monoid is of course a self-conjugate—see Definition 3.1—and thus an isomorphism. In fact, we have the following result.

**Lemma 5.2** *Involutive monoids (of the non-reversing kind) are ordinary monoids in the category of self-conjugates: the categories* $\mathbf{IMon}(\mathbf{C})$ *and* $\mathbf{Mon}(SC(\mathbf{C}))$ *are the same. Similarly in the commutative case,* $\mathbf{ICMon}(\mathbf{C}) = \mathbf{CMon}(SC(\mathbf{C}))$.

**Proof.** Since the tensors of $\mathbf{C}$ and $SC(\mathbf{C})$ coincide—see Proposition 4.4—we only need to check that the above definition precisely says that the unit $u$ and multiplication $m$ of an involutive monoid are maps in $SC(\mathbf{C})$ of the form:

$$\big(\overline{I} \xrightarrow{\zeta^{-1}} I\big) \xrightarrow{\;u\;} \big(\overline{M} \xrightarrow{j} M\big) \xleftarrow{\;m\;} \big(\overline{M} \xrightarrow{j} M\big) \otimes \big(\overline{M} \xrightarrow{j} M\big).$$

The unit $u$ is a map as indicated on the left if and only if $j \circ \overline{u} = u \circ \zeta^{-1}$. This is precisely the first square in Definition 5.1. Similarly, $m$ is map on the right if and only if $m \circ (j \otimes j) \circ \xi^{-1} = j \circ \overline{m}$. Again, this is exactly the (non-reversing) requirement in Definition 5.1. □

This lemma suggests a pattern for defining an involutive variant of certain categorical structure, namely by definiting this structure in the category of self-conjugates. Actions form an example, see Definition 5.5 below.

**Example 5.3** As we have observed before, the category $\mathbf{PoSets}$ of posets and monotone functions is involutive, via order-reversal $\overline{(X, \leq)} = (X, \geq)$. The poset $\mathbb{Z}$ of integers forms an involutive monoid in $\mathbf{PoSets}$, with minus $-\colon \overline{\mathbb{Z}} \to \mathbb{Z}$ as involution. Also, the positive rationals $\mathbb{Q}_{>0}$ or reals $\mathbb{R}_{>0}$ with multiplication $\cdot$, unit 1, and inverse $(-)^{-1}$ form involutive monoids in **PoSets**.

In the category **Cat** of categories, with finite products as monoidal structure, a monoid is a strictly monoidal category. If such a category $\mathbf{C}$ has a dagger $\dagger\colon \mathbf{C}^{\mathrm{op}} \to \mathbf{C}$ that commutes with these tensors (in the sense that $(f \otimes g)^\dagger = f^\dagger \otimes g^\dagger$, see *e.g.* [2]) then $\mathbf{C}$ is an involutive monoid in **Cat**.

Inside such a dagger symmetric (not necessarily strict) monoidal category $\mathbf{C}$ with dagger $(-)^\dagger\colon \mathbf{C}^{\mathrm{op}} \to \mathbf{C}$ the homset of scalars $I \to I$ is a commutative involutive monoid, with involution $s^- = s^\dagger$.

The tensor unit $I \in \mathbf{C}$ in an arbitrary involutive category $\mathbf{C}$ is a commutative involutive monoid object, with involution $\zeta^{-1}\colon \overline{I} \to I$.

We briefly describe free involutive monoids in the category **Sets** (with trivial involution), both of the reversing and non-reversing kind. We recall that the set $V^\star$ of finite lists $\langle v_1, \ldots, v_n\rangle$ of elements $v_i \in V$, is the free monoid on a set $V$, with empty list $\langle\rangle$ as unit and concatenation of lists as composition. We shall write 2 for the two-element set $2 = \{-, +\}$ of signs with negation (or involution) $-\colon 2 \to 2$ given by $-- = +$ and $-+ = -$.

**Proposition 5.4** *The free* non-reversing *involutive monoid on* $V \in$ **Sets** *is the set* $(2 \times V)^\star$ *of "signed" lists, with involution:*

$$\langle (b_1, v_1), \ldots, (b_n, v_n)\rangle^- = \langle (-b_1, v_1), \ldots, (-b_n, v_n)\rangle,$$

*where* $b_i \in 2$ *and* $v_i \in V$*. The free* reversing *involutive monoid als has* $(2 \times V)^\star$ *as carrier, but now with involution involving list reversal:*

$$\langle (b_1, v_1), \ldots, (b_n, v_n)\rangle^- = \langle (-b_n, v_n), \ldots, (-b_1, v_1)\rangle.$$

*In both cases we use* $\eta(v) = \langle (+, v)\rangle$ *as insertion* $\eta\colon V \to (2 \times V)^\star$.

**Proof.** Given an involutive monoid $M = (M, 1, \cdot, (-)^-)$ in **Sets**, a map $f\colon V \to M$ can be extended in a unique way to a map of non-reversing involutive monoids $\widehat{f}\colon (2 \times V)^\star \to M$, via

$$\widehat{f}\big(\langle (b_1, v_1), \ldots, (b_n, v_n)\rangle\big) = f(v_1)^{b_1} \cdot \ldots \cdot f(v_n)^{b_n},$$

where for $x \in M$ we write $x^+ = x$ and $x^-$ for the result of applying $M$'s involution $(-)^-$ to $x$. Clearly, $\widehat{f}$ preserves the unit and composition, and satisfies $\widehat{f} \circ \eta = f$. In the non-reversing case it preserves the involution:

$$\begin{aligned}
\widehat{f}\big(\langle (b_1, v_1), \ldots, (b_n, v_n)\rangle^-\big) &= \widehat{f}\langle (-b_1, v_1), \ldots, (-b_n, v_n)\rangle\big) \\
&= f(v_1)^{-b_1} \cdot \ldots \cdot f(v_n)^{-b_n} \\
&= \big(f(v_1)^{b_1}\big)^- \cdot \ldots \cdot \big(f(v_n)^{b_n}\big)^- \\
&= \big(f(v_1)^{b_1} \cdot \ldots \cdot f(v_n)^{b_n}\big)^- \\
&= \big(\widehat{f}(\langle (b_1, v_1), \ldots, (b_n, v_n)\rangle)\big)^-.
\end{aligned}$$

Similarly in the reversing case involution is preserved, because:

$$\begin{aligned} \widehat{f}\big(\langle(b_1, v_1), \ldots, (b_n, v_n)\rangle^{-}\big) &= \widehat{f}\langle(-b_n, v_n), \ldots, (-b_1, v_1)\rangle\big) \\ &= f(v_n)^{-b_n} \cdot \ldots \cdot f(v_1)^{-b_1} \\ &= \big(f(v_n)^{b_n}\big)^{-} \cdot \ldots \cdot \big(f(v_1)^{b_1}\big)^{-} \\ &= \big(f(v_1)^{b_1} \cdot \ldots \cdot f(v_n)^{b_n}\big)^{-} \\ &= \big(\widehat{f}(\langle(b_1, v_1), \ldots, (b_n, v_n)\rangle)\big)^{-}. \end{aligned}$$

□

For a non-reversing involutive monoid $M \in \mathbf{IMon}(\mathbf{C}) = \mathbf{Mon}(\mathit{SC}(\mathbf{C}))$ we can consider actions either in $\mathbf{C}$ or in $\mathit{SC}(\mathbf{C})$. The latter will be called 'involutive' actions.

**Definition 5.5** For an involutive monoid $M \in \mathbf{IMon}(\mathbf{C}) = \mathbf{Mon}(\mathit{SC}(\mathbf{C}))$ we write $\mathit{IAct}_M(\mathbf{C}) = \mathit{Act}_M(\mathit{SC}(\mathbf{C}))$ for the category of involutive actions. Its objects are actions in $\mathit{SC}(\mathbf{C})$ of the form:

$$\big(\overline{M} \xrightarrow{j} M\big) \otimes \big(\overline{X} \xrightarrow{j_X} X\big) \xrightarrow{a} \big(\overline{X} \xrightarrow{j_X} X\big) \qquad \text{i.e.} \qquad \begin{array}{ccc} \overline{M \otimes X} & \xrightarrow{\overline{a}} & \overline{X} \\ {\scriptstyle \xi^{-1}}\downarrow & & \\ \overline{M} \otimes \overline{X} & & \downarrow{\scriptstyle j_X} \\ {\scriptstyle j \otimes j_X}\downarrow & & \\ M \otimes X & \xrightarrow{a} & X \end{array}$$

together with the usual action requirements involving appropriate interaction with the unit and multiplication of the monoid $M$.

A morphism $f\colon (X, j_X) \to (Y, j_Y)$ in $\mathit{IAct}_M(\mathbf{C})$ is a morphism $f\colon X \to Y$ in $\mathbf{C}$ that is both a map of dualities, in $\mathit{SC}(\mathbf{C})$, and of actions, in $\mathit{Act}_M(\mathbf{C})$.

# 6 Involutions and algebras

This section introduces involutions on monads, and will focus on algebras of such monads. Familiarity with the basics of the theory of monads will be assumed, see *e.g.* [5,7,17,16]. Essentially, involutive monads are monads in the 2-category $\mathbf{ICat}$ of involutive categories. We describe them explicitly.

**Definition 6.1** Let $T = (T, \eta, \mu)$ be a monad on an involutive category $\mathbf{C}$. We shall call $T$ an involutive monad if $T\colon \mathbf{C} \to \mathbf{C}$ is an involutive functor, say via $\nu_X\colon T(\overline{X}) \to \overline{T(X)}$, and the unit $\eta$ and multiplication $\mu$ are involutive natural transformations. As a result, $\nu$ forms a distributive law of the monad $T$ over $\mathbf{C}$'s involution $\overline{(-)}$. This amounts to:

$$\begin{array}{ccc} \overline{X} & & \\ {\scriptstyle \eta}\downarrow & \searrow^{\overline{\eta}} & \\ T(\overline{X}) & \xrightarrow[\nu]{} & \overline{T(X)} \end{array} \qquad \begin{array}{ccccc} T^2(\overline{X}) & \xrightarrow{T(\nu)} & T(\overline{T(X)}) & \xrightarrow{\nu} & \overline{T^2(X)} \\ {\scriptstyle \mu}\downarrow & & & & \downarrow{\scriptstyle \overline{\mu}} \\ T(\overline{X}) & & \xrightarrow{\nu} & & \overline{T(X)} \end{array} \qquad \begin{array}{ccc} T(X) & \xrightarrow{\iota} & \overline{\overline{T(X)}} \\ {\scriptstyle T(\iota)}\downarrow & & \uparrow{\scriptstyle \overline{\nu}} \\ T(\overline{\overline{X}}) & \xrightarrow[\nu]{} & \overline{T(\overline{X})} \end{array}$$

This monad is called involutive (symmetric) monoidal if $T$ and $\eta, \mu$ are involutive (symmetric) monoidal.

With respect to the identity involution on a (symmetric monoidal) category $\mathbf{C}$, any monad is involutive via the identity distributive law. But the identity involution on a category may still give rise to meaningful involutive monads, as the semiring example below shows.

**Example 6.2** *(i)* Let $M = (M, m, u, j)$ be an involutive (non-reversing) monoid in an involutive category $\mathbf{C}$. As is well-known the functor $M \otimes (-) \colon \mathbf{C} \to \mathbf{C}$ is a monad; its unit and multiplication are:

$$X \xrightarrow[\cong]{\lambda^{-1}} I \otimes X \xrightarrow{u \otimes \mathrm{id}} M \otimes X \qquad M \otimes (M \otimes X) \xrightarrow[\cong]{\alpha} (M \otimes M) \otimes X \xrightarrow{m \otimes \mathrm{id}} M \otimes X.$$

Unsurprisingly, $M$'s involution $j$ makes this an involutive monad via:

$$\nu_X = \Big( M \otimes \overline{X} \xrightarrow[\cong]{j^{-1} \otimes \mathrm{id}} \overline{M} \otimes \overline{X} \xrightarrow[\cong]{\xi} \overline{M \otimes X} \Big).$$

*(ii)* Let $S$ be an involutive commutative semiring, *i.e.* a commutative semiring with an endomap $(-)^{-} \colon S \to S$ that is a semiring homomorphism with $s^{--} = s$. An obvious example is the set $\mathbb{C}$ of complex numbers with conjugation $\overline{a + ib} = a - ib$. Similarly, the Gaussian rational numbers (with $a, b \in \mathbb{Q}$ in $a+ib$) form an involutive semiring, albeit not a complete one.

Consider the multiset monad $\mathcal{M}_S \colon \mathbf{Sets} \to \mathbf{Sets}$ associated with $S$, where we use $\mathbf{Sets}$ as trivial involutive category, with the identity as involution. This monad is defined on a set $X$ as:

$$\mathcal{M}_S(X) = \{\varphi \colon X \to S \mid \mathit{supp}(\varphi) \text{ is finite}\}.$$

Such a multiset $\varphi \in \mathcal{M}_S(X)$ may be written as formal sum $s_1 x_1 + \cdots + s_k x_k$ where $\mathit{supp}(\varphi) = \{x_1, \ldots, x_k\}$ and $s_i = \varphi(x_i) \in S$ describes the "multiplicity" of the element $x_i \in X$. For more information, see *e.g.* [8]. The category of algebras of this monad is the category $\mathbf{Mod}_S$ of modules over $S$.

This monad is monoidal / commutative, because $S$ is commutative. It is involutive, with involution $\nu \colon \mathcal{M}_S(X) \to \mathcal{M}_S(X)$ given by $\nu(\sum_i s_i x_i) = \sum_i s_i^{-} x_i$.

For an involutive monad $T$ on an involutive category $\mathbf{C}$ we can consider two liftings, namely of the monad $T$ to self-dualities $\mathit{SC}(\mathbf{C})$ following Lemma 3.6, or of $\mathbf{C}$'s involution $\overline{(-)}$ to algebras $\mathit{Alg}(T)$, as in the following two diagrams.

$$\begin{array}{ccc} \mathit{SC}(\mathbf{C}) & \xrightarrow{\mathit{SC}(T)} & \mathit{SC}(\mathbf{C}) \\ \downarrow & & \downarrow \\ \mathbf{C} & \xrightarrow{T} & \mathbf{C} \end{array} \qquad\qquad \begin{array}{ccc} \mathit{Alg}(T) & \xrightarrow{\overline{(-)}} & \mathit{Alg}(T) \\ \downarrow & & \downarrow \\ \mathbf{C} & \xrightarrow{\overline{(-)}} & \mathbf{C} \end{array} \tag{8}$$

The lifting on the left yields a new monad $\mathit{SC}(T)$ because lifting in Lemma 3.6 is 2-functorial. The lifting on the right arises because $\nu$ in Definition 6.1 is a distributive law commuting with unit and multiplication. Explicitly, it is given by:

$$\overline{\big(T(X) \xrightarrow{a} X\big)} \stackrel{\text{def}}{=} \Big( T(\overline{X}) \xrightarrow{\nu_X} \overline{T(X)} \xrightarrow{\overline{a}} \overline{X} \Big). \tag{9}$$

We shall sometime refer to it as the 'conjugate' algebra, because conjugation of modules is an important instance, see Example 6.5 below.

**Proposition 6.3** *Suppose $T$ is an involutive monad on an involutive category $\mathbf{C}$. The category $\mathit{Alg}(T)$ is then also involutive via (9), and:*

(i) *$\mathit{Alg}(\mathcal{SC}(T)) = \mathcal{SC}(\mathit{Alg}(T))$, for which we sometimes write $\mathit{IAlg}(T)$;*

(ii) *the canonical adjunction $\mathit{Alg}(T) \leftrightarrows \mathbf{C}$ is an involutive one.*

**Proof.** Definition (9) yields a new algebra because $\nu$ is a distributive law. The involution $\mathrm{id}_{\mathit{Alg}(T)} \Rightarrow \overline{\overline{(-)}}$ on algebras is given by $\mathbf{C}$'s involution $\iota$, in:

$$\big(T(X) \xrightarrow{a} X\big) \xrightarrow[\cong]{\iota} \overline{\overline{\big(T(X) \xrightarrow{a} X\big)}}$$

It is not hard to see that $\iota$ is a map of algebras. The involution on a morphism $f$ of algebras is just $\overline{f}$.

For point (i) notice that on the one hand an $\mathcal{SC}(T)$-algebra is a map

$$\mathcal{SC}(T)\big(\overline{X} \xrightarrow{j_X} X\big) \xrightarrow{a} \big(\overline{X} \xrightarrow{j_X} X\big)$$

which is a $T$-algebra $a\colon T(X) \to X$ that is a map of self-conjugates, using (6) on the left in:

$$\begin{array}{ccc}
\overline{T(X)} & \xrightarrow{\overline{a}} & \overline{X} \\
{\scriptstyle \nu^{-1}}\downarrow & & \\
T(\overline{X}) & & \downarrow{\scriptstyle j_X} \\
{\scriptstyle T(j_X)}\downarrow & & \\
T(X) & \xrightarrow{a} & X
\end{array}$$

On the other hand a self-conjugate in $\mathit{Alg}(T)$ consists of an algebra $a$ with a map of the form:

$$\overline{\big(T(X) \xrightarrow{a} X\big)} \xrightarrow{j_X} \big(T(X) \xrightarrow{a} X\big)$$

which means that $j_X$ is a map of algebras:

$$\begin{array}{ccc}
T(\overline{X}) & \xrightarrow{T(j_X)} & T(X) \\
{\scriptstyle \nu}\downarrow & & \\
\overline{T(X)} & & \downarrow{\scriptstyle a} \\
{\scriptstyle \overline{a}}\downarrow & & \\
\overline{X} & \xrightarrow{j_X} & X
\end{array}$$

This is clearly the same as the previous rectangle.

As to the second point, the forgetful functor $\mathit{Alg}(T) \to \mathbf{C}$ clearly commutes with involution. The free functor $F\colon \mathbf{C} \to \mathit{Alg}(T)$, mapping $X$ to the algebra $\mu\colon T^2(X) \to T(X)$, is involutive via the map $F(\overline{X}) \to \overline{F(X)}$ that is simply $\nu$ itself by the (second) diagram in Definition 6.1, in:

$$\begin{array}{ccc} T^2(\overline{X}) & \xrightarrow{T(\nu)} & T(\overline{T(X)}) \\ & & \downarrow{\scriptstyle \nu} \\ \downarrow{\scriptstyle \mu_{\overline{X}}} & & \overline{T^2(X)} \\ & & \downarrow{\scriptstyle \overline{\mu}} \\ T(\overline{X}) & \xrightarrow{\nu} & \overline{T(X)} \end{array}$$ □

In a next step we would like to show that these categories of algebras of an involutive monoidal monad are also involutive monoidal categories. The monoidal structure is given by the standard construction of Anders Kock [15,14]. Tensors of algebras exist in case certain colimits exist. This is always the case with monads on sets, due to a result of Linton's, see [5, § 9.3, Prop. 4].

This tensor product $a \boxtimes b = (TX \xrightarrow{a} X) \boxtimes (TY \xrightarrow{b} Y)$ of algebras is such that algebra morphisms $a \boxtimes b \to c$ correspond to bimorphisms [15,14]. The latter can be defined abstractly. This tensor $a \boxtimes b$ arises as coequaliser in the category $\mathit{Alg}(T)$, of the form:

$$\begin{pmatrix} T^2(TX \otimes TY) \\ \downarrow{\scriptstyle \mu} \\ T(TX \otimes TY) \end{pmatrix} \underset{\mu \circ T(\xi)}{\overset{T(a \otimes b)}{\rightrightarrows}} \begin{pmatrix} T^2(X \otimes Y) \\ \downarrow{\scriptstyle \mu} \\ T(X \otimes Y) \end{pmatrix} \xrightarrow{t} \begin{pmatrix} T(X \boxtimes Y) \\ \downarrow{\scriptstyle a \boxtimes b} \\ X \boxtimes Y \end{pmatrix} \qquad (10)$$

We only give a sketch of the following result.

**Theorem 6.4** *Suppose $T$ is an involutive monoidal monad on an involutive monoidal category $\mathbf{C}$; assume the category $\mathit{Alg}(T)$ of algebras has enough coequalisers to make it monoidal, via (10). The category $\mathit{Alg}(T)$ is then also involutive monoidal, and the canonical adjunction $\mathit{Alg}(T) \leftrightarrows \mathbf{C}$ is an involutive monoidal one.*

*This result extends to* symmetric *monoidal structure, and also to closure (with exponents $\multimap$).*

**Proof.** For algebras $T(X) \xrightarrow{a} X$ and $T(Y) \xrightarrow{b} Y$ we need obtain a map of algebras $\xi^{\mathit{Alg}(T)} \colon \overline{a} \boxtimes \overline{b} \to \overline{a \boxtimes b}$ using the universal property described above. The map $\overline{\otimes} \circ \xi^{\mathbf{C}} \colon \overline{X} \otimes \overline{Y} \to \overline{X \otimes Y} \to \overline{X \boxtimes Y}$ is bilinear map, where $\otimes = t \circ \eta \colon X \otimes Y \to X \boxtimes Y$ is the universal bilinear map. Hence we obtain $\xi^{\mathit{Alg}(T)}$ with $\xi^{\mathit{Alg}(T)} \circ \otimes = \overline{\otimes} \circ \xi^{\mathbf{C}}$. The free algebra $F(I)$ is unit for the tensor $\boxtimes$ on $\mathit{Alg}(T)$ and comes with a map of algebras $\zeta^{\mathit{Alg}(T)} = \nu \circ T(\zeta^{\mathbf{C}}) \colon F(I) \to \overline{F(I)}$. □

**Example 6.5** In the context of Example 6.2 the construction (9) gives for an involutive commutative semiring $S$ an involution on the category $\mathbf{Mod}_S$ of $S$-modules, which maps a module $X$ to its conjugate space $\overline{X}$, with the same vectors but with scalar multiplication in $\overline{X}$ given by: $s \cdot_{\overline{X}} x = s^{-} \cdot_X x$, as we have already seen in (2).

Conjugate modules often occur in the context of Hilbert spaces. The category **Hilb** is indeed an involutive category, via this conjugation. Hence one can consider for instance involutive monoids in **Hilb**. They are sometimes called (unital) $H^*$-algebras.

We take a closer look at involutive monoids in categories of modules over an involutive semiring. They come close to the notion of $C^*$-algebra. Let $S$ be thus be an involutive semiring with the associated involutive category $\mathbf{Mod}_S$ of modules over $S$, like above. We shall write $\mathbf{IMod}_S$ for the associated category of *involutive* modules, which can be described in various ways:

$$\mathbf{IMod}_S = SC(\mathbf{Mod}_S) = SC(\mathcal{A}lg(\mathcal{M}_S)) = \mathcal{A}lg(SC(\mathcal{M}_S)) = I\mathcal{A}lg(\mathcal{M}_S).$$

An involutive module $M \in SC(\mathbf{Mod}_S)$ thus consists of a module $M = (M, +, 0, \cdot) \in \mathbf{Mod}_S$ together with an involution $(-)^- \colon \overline{M} \overset{\cong}{\Rightarrow} M$ in $\mathbf{Mod}_S$. This involution on $M$ preserves the monoid structure $(x+y)^- = x^- + y^-$ and $0^- = 0$, so that $M$ is an involutive monoid (in **Sets**). Its interaction with scalar multiplication is special, because of the conjugation $\overline{M}$ in its domain. It means that:

$$(s \cdot_{\overline{M}} x)^- = s \cdot_M x^- \qquad i.e. \qquad (s \cdot_M x)^- = s^- \cdot_M x^-. \tag{11}$$

A morphism $f \colon M \to N$ in $\mathbf{IMod}_S$ is a morphism of modules satisfying additionally $f(x^-) = f(x)^-$.

We add that the multiset monad $\mathcal{M}_S$ is 'additive', and so the products $\times$ in its category of algebras $\mathbf{Mod}_S$ are actually biproducts $\oplus$, see [8]. This additivity also holds for $SC(T)$, using Lemma 3.5, so that also $\mathbf{IMod}_S$ has biproducts $\oplus$. They are preserved by conjugation, essentially by Lemma 3.5.

# 7 The core of the GNS-construction

In this final section we wish to apply the theory developed so far to obtain what can be considered as the core of the (unital version of the) Gelfand-Naimark-Segal (GNS) construction [3], giving a bijective correspondence between states on $C^*$-algebras and certain sesquilinear maps. Roughly, for an involutive monoid $A$ in the category $\mathbf{IMod}_S$, as in Example 6.5, a state $f \colon A \to S$ gives rise to an inner product $\langle - \,|\, - \rangle \colon \overline{A} \otimes A \to S$ by $\langle a \,|\, b \rangle = f(a^- \cdot b)$, where $\cdot$ is the multiplication of the monoid $A$. Notice that using the involution $\overline{(-)}$ in the domain $\overline{A} \otimes A$ of the inner product gives a neat way of handling conjugation in the condition $\langle s \cdot a \,|\, b \rangle = s^- \cdot \langle a \,|\, b \rangle$, where this last $\cdot$ is the (scalar) multiplication of the semiring $S$ (which is the tensor unit in $\mathbf{Mod}_S$).

This induced inner product $\langle a \,|\, b \rangle = f(a^- \cdot b)$ satisfies two special properties that we capture abstractly below, namely: $\langle u \,|\, - \rangle = \langle - \,|\, u \rangle$ and $\langle a \cdot b \,|\, c \rangle = \langle a \,|\, b^- \cdot c \rangle$. These two properties appear as conditions (a) and (b) in the following result. Most commonly the inner product is described as a map $p \colon \overline{M} \otimes M \to I$ with the tensor unit as codomain, but the correspondence in the next result holds for an arbitrary self-conjugate $X$ instead of $I$.

**Theorem 7.1** *Let $M = (M, m, u, j)$ be a reversing involutive monoid in an involutive symmetric monoidal category (ISMC) $\mathbf{C}$ and let $j_X \colon \overline{X} \to X$ be a self-conjugate. Consider the following two properties of a map $p \colon \overline{M} \otimes M \to X$.*

(a) *Sameness when restricted to units:*

$$\begin{array}{ccccccc}
\overline{M} & \xrightarrow[\cong]{\rho^{-1}} & \overline{M}\otimes I & \xrightarrow{\mathrm{id}\otimes u} & \overline{M}\otimes M & \xrightarrow{p} & X \\
{\scriptstyle j}\downarrow & & & & & & \uparrow{\scriptstyle p} \\
M & \xrightarrow[\cong]{\lambda^{-1}} & I\otimes M & \xrightarrow{\zeta\otimes\mathrm{id}} & \overline{I}\otimes M & \xrightarrow{\overline{u}\otimes\mathrm{id}} & \overline{M}\otimes M
\end{array}$$

(b) *Shifting of multiplications:*

$$\begin{array}{ccccccccc}
(\overline{M}\otimes\overline{M})\otimes M & \xrightarrow{\xi\otimes\mathrm{id}} & \overline{(M\otimes M)}\otimes M & & \xrightarrow{\overline{m}\otimes\mathrm{id}} & & \overline{M}\otimes M & \xrightarrow{p} & X \\
{\scriptstyle \gamma\otimes\mathrm{id}}\downarrow\cong & & & & & & & & \uparrow{\scriptstyle p} \\
(\overline{M}\otimes\overline{M})\otimes M & \xrightarrow[\cong]{\alpha^{-1}} & \overline{M}\otimes(\overline{M}\otimes M) & \xrightarrow{\mathrm{id}\otimes(j\otimes\mathrm{id})} & \overline{M}\otimes(M\otimes M) & & \xrightarrow{\mathrm{id}\otimes m} & & \overline{M}\otimes M
\end{array}$$

*Then there is a bijective correspondence between maps in* $SC(\mathbf{C})$,

$$\frac{M \xrightarrow{f} X}{\overline{M}\otimes M \xrightarrow[p]{} X \ \ \textit{satisfying (a) and (b)}} \qquad (12)$$

*where* $\overline{M}\otimes M$ *is provided with the "twist" conjugate* $t$ *defined as:*

$$t \stackrel{def}{=} \Big(\overline{\overline{M}\otimes M} \xrightarrow{\overline{\mathrm{id}\otimes\iota_M}} \overline{\overline{M}\otimes\overline{\overline{M}}} \xrightarrow{\overline{\xi}} \overline{\overline{M\otimes\overline{M}}} \xrightarrow{\iota^{-1}} M\otimes\overline{M} \xrightarrow{\gamma} \overline{M}\otimes M\Big).$$

**Proof.** Verification of this correspondence involves many details, of which we present the essentials. Given $f\colon M\to X$ in $SC(\mathbf{C})$, we define

$$\widehat{f} \stackrel{\text{def}}{=} \Big(\overline{M}\otimes M \xrightarrow{j\otimes\mathrm{id}} M\otimes M \xrightarrow{m} M \xrightarrow{f} X\Big).$$

This $\widehat{f}$ is a map in $SC(\mathbf{C})$ since we have $\widehat{f}\circ t = j_X\circ\overline{\widehat{f}}$ in:

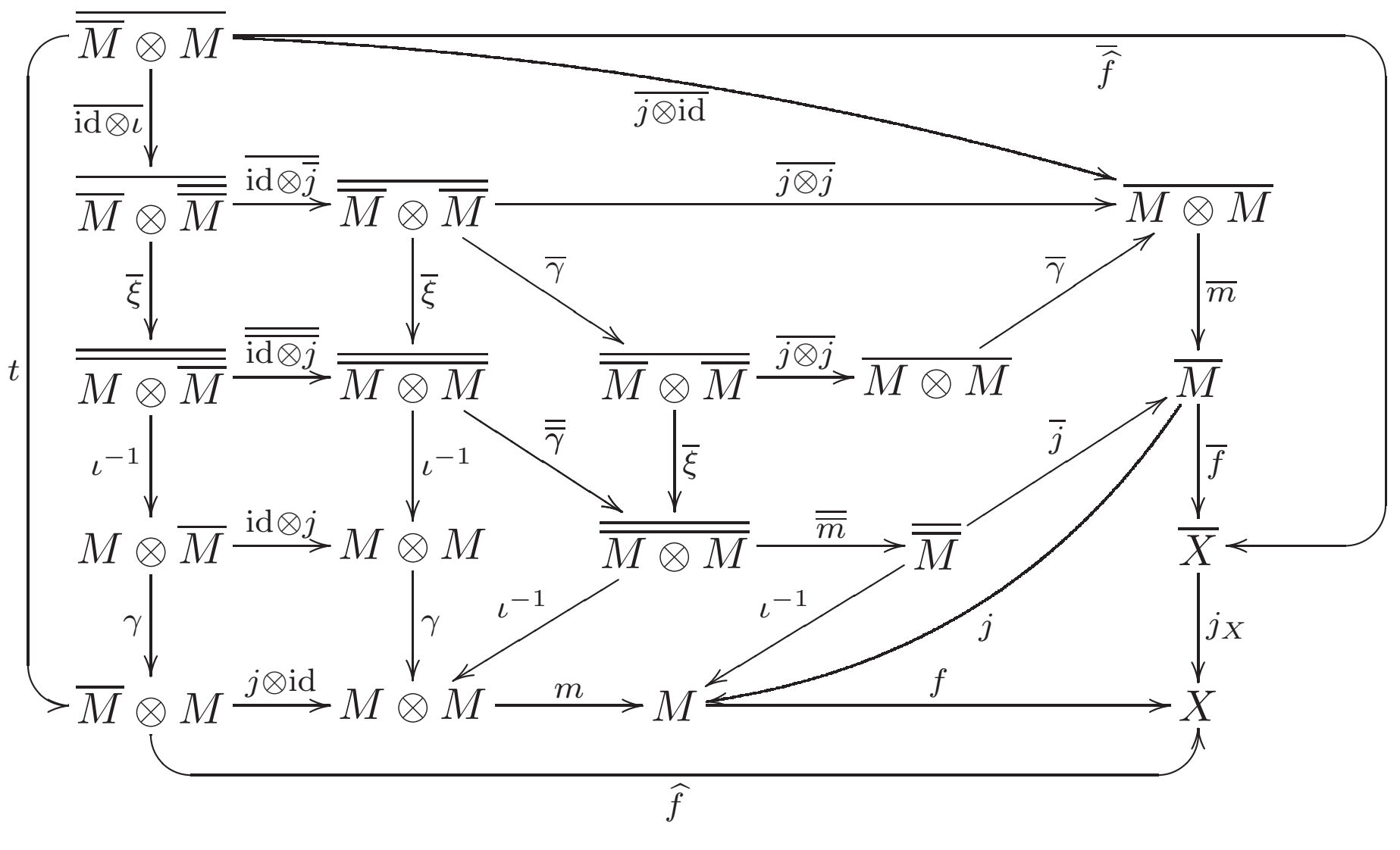

It is not hard to show that $\widehat{f}$ satisfies the above two properties (a) and (b).

Conversely, given $p\colon M \otimes \overline{M} \to X$ in $SC(\mathbf{C})$ we take:

$$\widehat{p} = \Big( M \xrightarrow{\lambda^{-1}} I \otimes M \xrightarrow{\zeta \otimes \mathrm{id}} \overline{I} \otimes M \xrightarrow{\overline{e} \otimes \mathrm{id}} \overline{M} \otimes M \xrightarrow{p} X \Big).$$

Using property (a) one shows that $\widehat{p}$ is a map of self-dualities. Next we check that we get a bijective correspondence (12). Starting from $f\colon M \to X$ we get $\widehat{\widehat{f}} = f$ in:

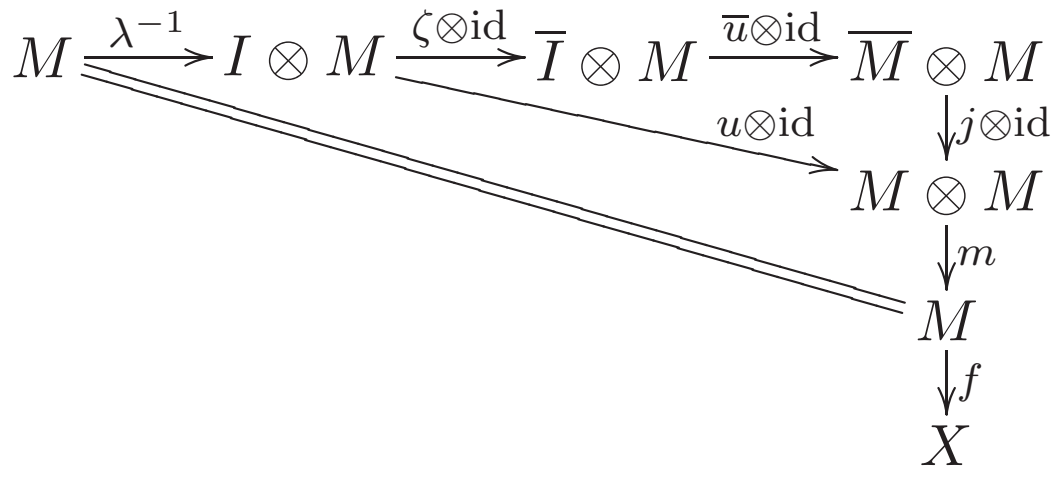

The verification that $\widehat{\widehat{p}} = p$ for $p\colon \overline{M} \otimes M \to X$ is more involved and requires property $(b)$, see:

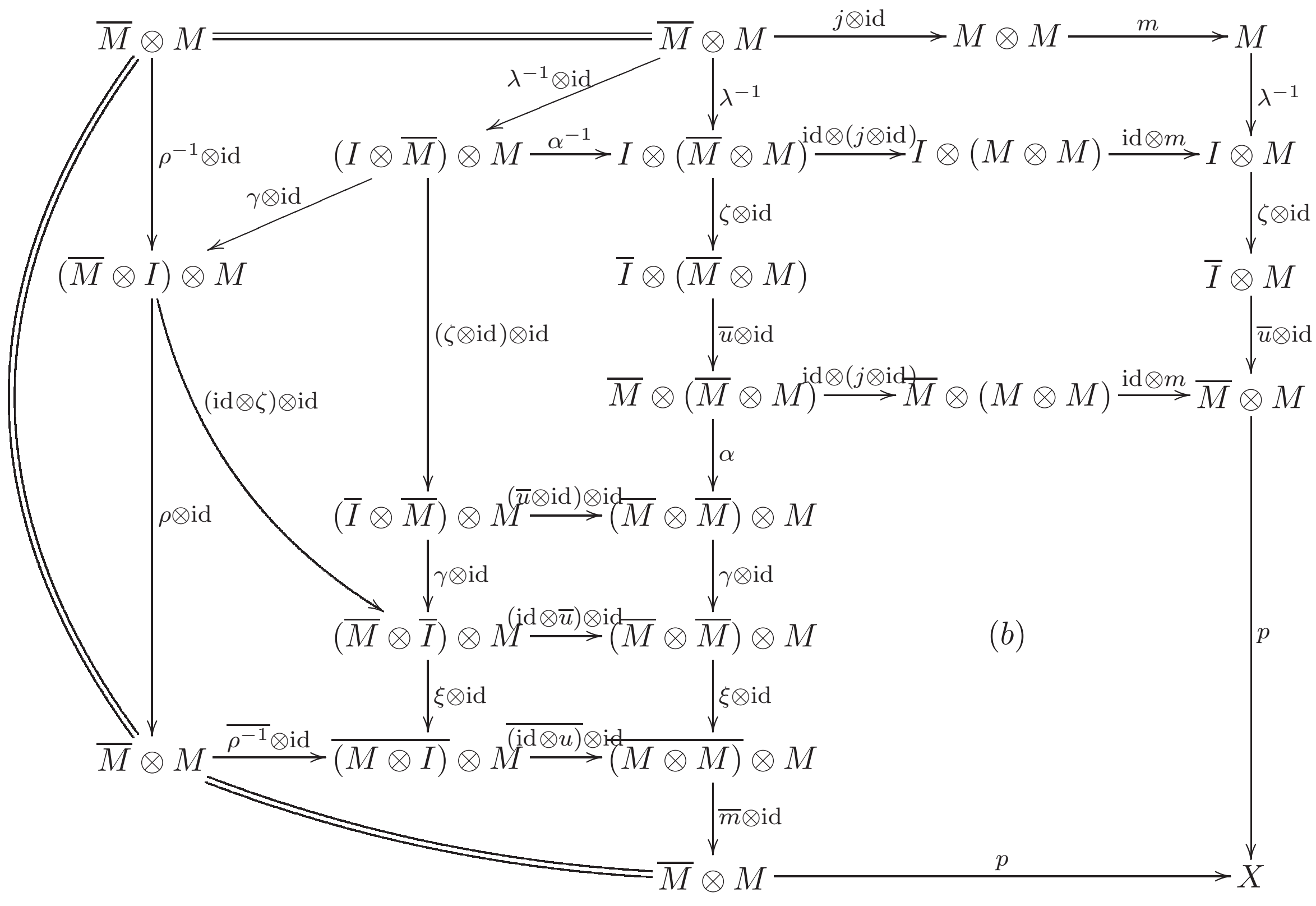

Remaining details are left to the reader. □

As said, this result only captures the heart of the GNS construction [3]. The whole construction additionally involves suitable quotients, in order to identify points $a, b$ with $\langle a \,|\, b \rangle = 0$, and completions, in order to get a complete metric space, and thus a Hilbert space.

*Acknowledgements*

Thanks to Chris Heunen and Jorik Mandemaker for useful feedback.